\documentclass[accepted]{uai2021} 

\usepackage[british]{babel}

\usepackage{natbib} 
\usepackage{mathtools} 
\usepackage{booktabs} 
\usepackage{tikz} 
\usepackage{algorithm2e}
\usepackage{multirow}

\usepackage{listings}
\usetikzlibrary{calc}
\usepackage[many]{tcolorbox}
\tcbuselibrary{listings}

\title{PASOCS: A Parallel Approximate Solver for Probabilistic Logic Programs under the Credal Semantics}

\author[1]{\href{mailto:David Tuckey <dwt17@ic.ac.uk>?Subject=Your UAI 2021 paper}{David Tuckey}{}} 
\author[2]{Alessandra Russo}
\author[2]{Krysia Broda}
\affil[1,2]{%
    Department of Computing\\
    Imperial College London\\
    London, UK
}
\affil{%
    dwt17@ic.ac.uk
}

\begin{document}
\maketitle

\begin{abstract}
The Credal semantics is a probabilistic extension of the answer set semantics which can be applied to programs that may or may not be stratified. It assigns to atoms a set of acceptable probability distributions characterised by its lower and upper bounds. Performing exact probabilistic inference in the Credal semantics is computationally intractable. This paper presents a first solver, based on sampling, for probabilistic inference under the Credal semantics called PASOCS (Parallel Approximate SOlver for the Credal Semantics). PASOCS performs both exact and approximate inference for queries given evidence. Approximate solutions can be generated using any of the following sampling methods: naive sampling, Metropolis-Hastings and Gibbs Markov Chain Monte-Carlo. We evaluate the fidelity and performance of our system when applied to both stratified and non-stratified programs. We perform a sanity check by comparing PASOCS to available systems for stratified programs, where the semantics agree, and show that our system is competitive on unstratified programs.
\end{abstract}


\section{Introduction}\label{sec:intro}
Probabilistic logic programming (PLP) is a subfield of Artificial Intelligence aimed at handling uncertainty in formal arguments by combining logic programming with probability theory~\citep{Riguzzi2018}. Traditional logic programming languages are extended with constructs, such as probabilistic facts and annotated disjunctions, to model complex relations over probabilistic outcomes. Different probabilistic semantics for PLP have been proposed in the literature of which Sato's distribution semantics \citep{Sato1995} is to date the most prominent one. It underpins a variety of PLP systems including Problog \citep{Fierens2015}, PRISM \citep{Sato2001} and Cplint \citep{Riguzzi2007}, which extend Prolog-based logic programming with probabilistic inference. Probabilistic extensions have also been proposed within the context of Answer Set Programming (ASP), leading to systems such as P-log \citep{Baral2004}, $\mathrm{LP}^{\mathrm{MLN}}$ \citep{Lee2015} and PrASP \citep{Nickles2015}. Although all these approaches and systems make use of different methodologies ranging from  knowledge compilation \citep{Fierens2015} to various forms of sampling \citep{Nampally2014,Azzolini2020,Shterionov2010} to compute probabilistic outcomes, they all share the characteristic of computing point probabilities, assuming a single probability distribution over the outcomes of a PLP. 

In real-world applications uncertainty comes not only in the form of likelihood of truth or falsity of Boolean random variables, but also as a result of incomplete knowledge or non-deterministic choices, which are not quantifiable probabilistically. These concepts are expressible in Answer Set Programming, for instance through negative loops and choice rules, and semantically lead to multiple models (answer sets \citep{Gelfond1988}) for a given answer set program. The probabilistic inference has, in this case, to cater for the possibility of multiple answer sets for a given total choice of probabilistic variables. The credal semantics captures this property and generalises Sato’s distribution semantics \citep{Lukasiewicz2005, Cozman2020}. It attributes to the elements in a probabilistic logic program a set of probability measures instead of single point values. This allows incomplete knowledge to be represented on top of stochastic behaviours  \citep{Halpern2018a} without having to make any stochastic assumptions on missing information. The result is that each element in the program has a lower and upper bound probability akin to worst and best case scenarios. To the best of our knowledge no probabilistic inference system has been proposed for the Credal semantics. 

This paper addresses this problem by presenting the first approximate solver for the Credal semantics, named PASOCS (Parallel Approximate SOlver for the Credal Semantics)\footnote{\href{https://github.com/David-Tuc/pasocs}{https://github.com/David-Tuc/pasocs}}.  It is applicable to PLPs that are stratified and for which the Credal semantics collapses to the point probability of Sato’s distribution semantics. But for more general problems, where the logic programs involved have multiple answer sets, PASOCS returns probability bounds as described by the Credal semantics. Performing probabilistic inference under the Credal semantics is inherently a hard computational task \citep{Maua2020}. Though our solver supports exact inference through the computation of every possible world, its true aim is to allow for approximate inference by sampling the space of possible worlds and using Clingo \citep{Gebser2014} to solve the resulting logic programs. The system incorporates different sampling methods: naive sampling, Metropolis-Hastings and Gibbs Markov Chain Monte Carlo (MCMC) sampling. Solving calls are made in parallel to leverage the availability of multi-core processing nodes and scale to high-throughput computing clusters. 

We evaluate the fidelity and performance of our system when applied to both stratified and non-stratified programs. We perform a sanity check by comparing PASOCS to available PLP systems for stratified programs, where the different semantics agree, and show that our system is competitive while working with a less restrictive semantics on unstratified programs.

The paper is structured as follows: Section \ref{sec::background} introduces the credal semantics to allow for the presentation of PASOCS in Section \ref{sec::PASOCS}. We evaluate our system in Section \ref{sec::experiments}. We then present related works in Section \ref{sec::related_works} and conclude in Section \ref{sec::conclusion}.

\section{Background}
\label{sec::background}

This section introduces the main two notions used throughout the paper: the syntax and semantics for Answer Set Programming (ASP)~\citep{Gelfond1988}, and the Credal semantics \citep{Lukasiewicz2005, Cozman2020} for PLP. 

\subsection{Answer set semantics}

We assume the following subset of the ASP language\footnote{For the full syntax and semantics of ASP please see~\citep{Gebser2014}.}. A {\em literal} can be either an atom of the form $p(t_1,...,t_n)$, where $p$ is a predicate of arity $n$ and $t_i$ are either constants or variables, or its {\em default negation} $not\textnormal{ }p(t_1,...,t_n)$, where $not$ is {\em negation as failure}. A rule $r$ is of the form $A_1\leftarrow B_1, ..., B_n,not\textnormal{ }B_{n+1}, ..., not\textnormal{ }B_m$ where all $A_1$ and $B_{i}$ are atoms. We call $A_1$ the {\em head} of the rule (denoted $h(r)$) and 
$B_1, ..., B_n,not\textnormal{ }B_{n+1}, ..., not\textnormal{ }B_m$ the {\em body} of the rule. Specifically, we refer to
$\{B_1,...,B_n\}$ as the {\em positive body} literals of the rule $r$ (denoted $B^+(r)$) and to  $\{not\textnormal{ }B_{n+1},...,not\textnormal{ }B_m\}$ as the {\em negative body} literals of the rule $r$ (denoted $B^-(r)$). A rule $r$ is said to be {\em definite} if $B^-(r)=\emptyset$. A
{\em fact} is a rule with an empty body. An ASP program is a finite set of rules.
Given an ASP program $P$, the Herbrand Base of $P$, denoted as $HB_{P}$, is the set of all ground (variable free) atoms that can be formed from the predicates and constants that appear in $P$. The grounding of an ASP program $P$, denoted $gr(P)$, is the program composed of all possible ground rules constructed from the rules in $P$ by substituting variables with constants that appear in $P$.  
Given an ASP program $P$ and a set $I\subseteq HB_P$, the {\em reduct} of $P$ with respect to $I$, denoted ($P^{I}$), is the ground program constructed from $gr(P)$ by removing any rule whose body contains a negative body literal $not\textnormal{ }B_j$ where $B_j\in I$, and removing any negative literal in the remaining rules. Note that $P^{I}$ is, by construction, a ground definite program. i.e. composed only of definite rules and facts. A model of $P^{I}$ is an interpretation $I^{'}\subseteq HB_P$ (a set of ground atoms) such that for each rule $r \in P^I$, $h(r) \in I'$ or $B^+(r) \not\subset I'$. The Least Herbrand Model of $P^{I}$, denoted as $LHM(P^{I})$, is the minimal model (with respect to set inclusion) of $P^{I}$. An interpretation $A\subseteq HB_P$ is an {\em Answer Set} of an ASP program $P$  if and only if $A=LHM(P^{A})$. The set of Answer Sets of a program $P$ is denoted  $AS(P)$.

An ASP program $P$ is \textit{stratified} if it can be written as $P = P_1 \cup P_2 \cup ... \cup P_k$ such that $P_i\cap P_j=\emptyset$ for any $1\leq i,j\leq k$, any positive body literal $B_i$ of a rule in $P_i$ is the head of a rule in $P_j$, for $j \leq i$, and the atom of any negative body literal $not\textnormal{ }B_j$ of a rule in $P_i$ is the head of a rule in $P_j$ with $j<i$. 
A stratified ASP program $P$ has only one Answer Set $A$ (i.e., $|AS(P)|=1$), which is the only interpretation such as $A=LHM(P^{A})$. Non-stratified ASP programs may have multiple Answer Sets (i.e., $|AS(P)|> 1$). 

\subsection{Credal semantics}
\label{sec::cs}

\begin{figure}[t]
    \centering
     \begin{lstlisting}[]
        0.3::a.
        p :- not q, a.
        q :- not p.
          \end{lstlisting}
    \caption{Example of simplistic probabilistic logic program. $a$ is true with probability $0.3$ and false with probability $0.7$. This program has only two total choices: $\{a\}$ and $\{\}$. With the total choice where $a$ is false, there is only one answer set $\{q\}$, and when $a$ is true there are two possible answer sets $\{p, a\}$ and $\{q, a\}$. This makes that $\underline{P}(q) = 0.7$, $\overline{P}(q) = 1$, $\underline{P}(p) = 0$ and $\overline{P}(p) = 0.3$. }
    \label{fig::ex}
\end{figure}

The Credal semantics is a generalisation of Sato's distribution semantics \citep{Sato1995} which allows for probabilistic logic programs to be non stratified. In this paper, we consider PLP programs $P=<P_l, P_f>$ that are composed of a (possibly non-stratified) logic program $P_l$ and a set $P_f$ of {\em probabilistic facts} of the form $pr::B$ where $B$ is a ground atom, representing a Boolean random variable, and $pr\in[0,1]$ defines its probability distribution: $p(B =\mbox{true}) = pr$  and $p(B=\mbox{false})=1-pr$. Informally, $B$ is true with probability $pr$ and false with probability $1-pr$. Probabilistic facts are assumed to be independent Boolean random variables and to have unique ground atoms (i.e., an atom $B$ of a probabilistic fact may appear only once in $P_f$). PLP programs $P=<P_l, P_f>$ are also assumed to satisfy the {\em disjoint condition}, that is atoms of probabilistic facts cannot appear as head atoms of any rule in $gr(P_l)$. 
Atoms of probabilistic facts can be ``chosen" to be true or false. This choice can be seen as a probabilistic outcome of their truth value governed by their probability distribution. We refer to a {\em total choice} $C\in 2^{P_f}$ as a subset of probabilistic facts that are chosen to be true: $B$ is considered to be true in a total choice $C$ if $(pr::B)\in C$. Each choice $C$ has an associated probability given by 
$Prob(C) = \prod_{(pr::B_i) \in C}\; pr\;\prod_{(pr::B_j) \not\in C}\;(1-pr)$, since probabilistic facts are assumed to be independent. Given a total choice $C$, we denote with $C_p =\{B|(pr::B)\in C\}$ the set of ground atoms that appear in $C$ and with $\bigwedge C =\bigwedge_{(pr::B)\in C}\textnormal{ } B$  the logical conjunction of the atoms chosen to be true in $C$. Given a PLP $P=<P_l, P_f>$, we define the Herbrand Base as $HB_{P} = HB_{P_l}\cup\{B|(pr::B) \in P_f\}$. 

For a PLP $P$, probability measures $Pr$ are defined over its interpretations $I\subseteq HB_P$. 
Given an atom $B$ in the language of a PLP program $P$, we can define its probability (by surcharge of notation) as $Pr(B) = \sum_{I\subseteq HB_P, \;I\models B} Pr(I)$. We also define the probability of a conjunctive formula $F$ as $Pr(F) = \sum_{I\subseteq HB_P,\;I\models F} Pr(I)$, the sum of the probabilities of the interpretations in which $F$ is true. A PLP is said to be consistent under the Credal semantics if for all total choices $C$ we have $AS(P_l \cup C_p) \neq \emptyset$. The Credal semantics links the notion of probability of a total choice with the notion of answer sets \citep{Cozman2017}. In particular, given a PLP $P$, a \textit{probability model} $Pr$ is a probability measure such that every interpretation $I\subseteq HB_P$ with $Pr(I)>0$ is an answer set of $P_l\cup C_p$ for some total choice $C$ ($Pr(I)>0\Rightarrow\exists C \in 2^{P_f} \textnormal{ s.t. }I\in AS(P\cup C_p)$), and for all total choices C, $Pr(\bigwedge C) = Prob(C)$.

The Credal semantics of a PLP program $P=\langle P_l, P
_f\rangle$ is its set of probability models. Given a query with evidence $q = (Q|E)$ where $Q$ and $E$ are sets of truth assignments to ground atoms in $HB_P$, the Credal semantics associates to it a set of probability measures, to which we can give its upper bound $\overline{P}(Q|E)$ and lower bound $\underline{P}(Q|E)$. An example is given in Figure \ref{fig::ex}. \citet{Cozman2017} provide an algorithm to compute the upper and lower bound of a given query with evidence, which we report in Algorithm \ref{alg::cs}. We say that a PLP $P$ is stratified when $P_l$ is stratified. For such programs, $\underline{P}(Q|E) = \overline{P}(Q|E)$ for any query $(Q|E)$.

\begin{algorithm}[t]
        \SetAlgoLined
        \KwData{PLP $P=<P_l, P_f>$ and query $(Q|E)$}
        \KwResult{$[\underline{P}(Q|E),\overline{P}(Q|E)]$}
        a,b,c,d = 0\;
        \ForEach{Total choice C}{
        \lIf{$Q\cup E$ is true in every answer sets $AS(P\cup C_p)$}{
        a $\leftarrow$ a + $Prob(C)$}
        \lIf{$Q\cup E$ is true in some answer sets $AS(P\cup C_p)$}{
        b $\leftarrow$ b + $Prob(C)$}
        \lIf{$Q$ is false and $E$ is true in every answer sets $AS(P\cup C_p)$}{
        c $\leftarrow$ c + $Prob(C)$}
        \lIf{$Q$ is false and $E$ is true in some answer sets $AS(P\cup C_p)$}{
        d $\leftarrow$ d + $Prob(C)$}
        }
        \uIf{$ b+c=0 $ and $ d>0 $}{\KwRet [0,0]}
        \uElseIf{$ a+d=0 $ and $ b>0 $}{\KwRet [1,1]}
        \Else{\KwRet $[a/(a+d), b/(b+c)]$}
        \caption{Algorithm to compute the lower and upper bound of a query from answer sets. From \citep{Cozman2017}}
        \label{alg::cs}
\end{algorithm}

\subsection{Sampling}

Sampling is a process by which one randomly selects a subset of the population to estimate a distribution \citep{Montgomery1994}. It has been applied to different probabilistic logic programming settings under various semantics \citep{Shterionov2010,Nickles2015,Azzolini2019,Nampally2014}.

%
%
A well known method for sampling uses the Markov Chain Monte-Carlo method where the next sample is built from the existing sample, instead of being drawn independently from the distribution. In this paper, we will use the well known Metropolis-Hastings (MH) \citep{Hastings1970} and Gibbs \citep{Geman1984} algorithms, which are both MCMC methods. The MH MCMC method consists of creating the samples using an intermediate Markov chain, which is easier to traverse, and accepting these new samples with a certain ``acceptance'' probability. Gibbs MCMC sampling consists of re-sampling one (or a fixed amount $k$ in the case of block Gibbs sampling) of the random variables involved using their respective distribution at each new sample $n+1$. It is a special case of MH sampling where the acceptance probability is always $1$. For both methods, it is common practice to perform a \textit{burning} step, which means taking $b$ sampling steps at the beginning and discard them, in order to minimize the impact of the initialization.

\section{PASOCS}
\label{sec::PASOCS}

We now present our solver PASOCS (Parallel Approximate SOlver for the Credal Semantics), based on sampling, for performing probabilistic inference under the Credal semantics. PASOCS is a system aimed at computing the lower and upper bounds of queries with evidence for probabilistic ASP programs. The system allows for exact and approximate inference by computing, or sampling, the set of total choices and solving the resulting ASP program using the ASP solver Clingo \citep{Gebser2008}. The calls to Clingo are parallelized, which allows scaling the computation over multiple CPUs. In what follows, we present the input probabilistic ASP language accepted by PASOCS and give details of the the sampling parameters. 

\subsection{Input Language}
\label{sec::input}

A PASOCS program $P=P_{ASP} \cup P_A$ is an ASP program ($P_{ASP}$) extended with a set of rules ($P_A$) annotated with probabilities. Annotated rules are of the following form, where $p$, $p1$, $p2$, $p3$ are ground atoms and $L_1...L_m$ are literals. 

\vspace{-3mm}
\begin{equation*}
    \begin{split}
        &pr::p. \textnormal{ (1)} \\
        &pr::p \textnormal{ :- }L_1, ..., L_m. \textnormal{ (2)} \\
        &pr1::p1;pr2::p2;pr3::p3. \textnormal{ (3)}\\
        &pr1::p1;pr2::p2;pr3::p3\textnormal{ :- }L_1, ..., L_m. \textnormal{ (4)}
    \end{split}
\end{equation*}
A probabilistic ASP program $P=P_{ASP} \cup P_A$ can be translated into an equivalent PLP program $<P_l, P_f>$. This is done as follows. Firstly, we initialise $P_l=P_{ASP}$. Annotated rule (1) is already a probabilistic fact, so we add $pr::p$ to $P_f$. Annotated rule (2) is a syntactic sugar for a "probabilistic clause". We add a probabilistic fact $pr::p_i$ to $P_f$ and add the clause $p :- L_1, ..., L_m, p_i$ to $P_l$, where $i$ is a unique identifier. Annotated rule (3) is an annotated disjunction, meaning that a "probabilistic fact" takes value $p1$ with probability $pr1$, value $p2$ with probability $pr2$, etc... The sum of the probabilities in the annotated disjunction has to be less than or equal to 1. This annotated rule (3) is transformed, using the mechanism described in \citep[Chapter~3]{Gutmann2011a}, into a set of clauses, which are added to $P_l$, and related probabilistic facts, which are added to $P_f$ (see Figure \ref{fig::disjunction} for an example translation). 
Annotated rules of type (4) are translated similarly to the rule of type (3) with the addition that the body elements $L_1,....L_m$ are added to the body of all the rules created from the disjunctive head.

\begin{figure}[t]
    \centering
     \begin{lstlisting}[]
        a::pf1.
        b::pf2.
        c::pf3.
        p1 :- pf1.
        p2 :- pf2, not pf1.
        p3 :- pf3, not pf1, not pf2.
    \end{lstlisting}
    \caption{Translation of annotated rule (3) in Section \ref{sec::input} following the method in \citep[Chapter~3]{Gutmann2011a}. The probabilistic facts have probabilities $a=pr1$, $b = \dfrac{pr2}{1-pr1}$ and $c = \dfrac{pr3}{1-pr1-pr2}$}
    \label{fig::disjunction}
\end{figure}

Queries can be written as part of the program and are expressed using the $\#query$ keyword. An example query would be:
\begin{equation*}
    \#query(p1,not p2, p3|p4:true, p5:false,p6:true).
\end{equation*}

\vspace{-3mm}
where $p1, p2, p3, p4, p5$ and $p6$ are ground atoms. $p1,not p2, p3$ is the query and $p4:true, p5:false,p6:true$ is the provided evidence. In this example, we want to know the probability of $p1$ and $p3$ to be true and $p2$ to be false (at the same time), given that $p4$ and $p6$ are true and $p5$ is false.
In PASOCS, we can specify multiple queries at the same time, which can all have a different set of evidence. One advantage of PASOCS is that it can run all the queries at the same time and does not require multiple calls to answer queries with different evidence. 

\subsection{Exact Inference}

The PASOCS system performs exact inference for a given set of queries in three steps: firstly it computes the set of all total choices from the transformed program $<P_l, P_f>$; secondly it generates the answer sets corresponding to each of these total choices; and finally it uses Algorithm \ref{alg::cs} to obtain the lower and upper bounds probabilities. We have evaluated PASOCS's exact inference and compared it with other existing methods in the case of stratified PLP programs. Results, given in Section~\ref{sec::experiments}, show that the computational time is, as expected, exponential in the number of probabilistic variables (see Table~\ref{Table1}). This is because given $n$ probabilistic facts there are $2^n$ total choices. To address this problem PASOCS uses sampling for approximate solving of probabilistic inference tasks.

\subsection{Approximate solving}
PASOCS performs approximate solving through sampling total choices. Specifically, PASOCS makes use of three different sampling algorithms: Naive Sampling, Metropolis-Hasting MCMC and Gibbs MCMC. These three sampling methods share the same stopping criteria, which are user defined. Total choices are sampled and queries are evaluated on the answer sets resulting from each of the sampled total choices. All queries are evaluated simultaneously on the same samples. If a query has evidence, PASOCS instructs it to ignore a sample when the evidence is not true in any of the resulting answer sets. This means that different queries, even if evaluated on the same set of samples, may not have the same count of number of samples. To evaluate the upper and lower bound of a query, PASOCS uses Algorithm \ref{alg::cs} where the probability of a total choice, $Prob(c)$, is replaced by $1$, and the resulting values $a,b,c$ and $d$ are divided by the number of samples that the query has counted. \\

PASOCS considers the predicted lower and upper bounds of a query separately and so computes the uncertainty on both independently. With each bound it estimates following a Bernoulli distribution, we formulate the uncertainty as: \citep{Montgomery1994}:
\begin{equation}
    U = 2*perc*\sqrt{\dfrac{p(1-p)}{N}}
\end{equation}
where $p$ is the estimated lower or upper bound for the query, $N$ is the number of samples the query has counted and $perc$ is a user defined parameter that is the percentile to use (by default 1.96 for the 95\% confidence bounds). The system stops sampling when the uncertainty of the lower and upper bounds (computed separately) for all queries is under a certain \textit{threshold} defined by the user. For $p=0$ or $p=1$, the uncertainty is $0$. In these cases the system does not consider the user defined threshold but instead continues sampling until a minimum amount of sample $min_{sample}$ has been counted for each query. Finally, a  user defined parameter $max_{sample}$ provides the system with the maximum number of samples that it is allowed to take in total. The rest of this section describes the use of the three different sampling methods. 

\paragraph{Naive Sampling.}
The naive sampling algorithm samples total choices by independently sampling each probabilistic fact using their associated probabilities in $P_f$. Each sample is independent from the next.

\paragraph{Metropolis-Hastings Sampling.}
For the MH MCMC algorithm, PASOCS randomly initialises the value for each probabilistic fact and performs a user defined number of burning steps. To build a sample $n+1$ from an existing sample $n$, PASOCS switches the value of each probabilistic fact with probability $p_{change}$ (user defined, default is $0.3$). This makes that the acceptance probability for sample $S_{n+1}$ is $max(1, \dfrac{Prob(S_{n+1})}{Prob(S_n)})$ where $S_{n+1}$ is the sample $n+1$ and $S_n$ is the sample $n$, and $Prob$ being the probability of the total choices as defined in Section \ref{sec::cs}. 

\paragraph{Gibbs sampling.} 
PASOCS can also use Gibbs MCMC sampling, in particular  block Gibbs MCMC sampling. In this case the size $bk$ of the block is user defined. The system initialises the values of the probabilistic facts randomly and performs a burn phase at the beginning of the sampling. Then at  each iteration, $bk$ probabilistic facts are sampled at a time using their respective probabilities defined in $P_f$ while keeping the others fixed. By default $bk=1$. 

\section{Experiments}
\label{sec::experiments}

We demonstrate the capabilities of PASOCS on two tasks to evaluate the sampling mechanism. We compare against Problog, Cplint (PITA \citep{Riguzzi2011} and MCINTYRE \citep{Riguzzi2011b}), Diff-Sat and  $\mathrm{LP}^{\mathrm{MLN}}$ both quantitatively and qualitatively. We report running times of all the systems when relevant. We ran all the systems (apart from Cplint\footnote{Which we ran from http://cplint.eu}) on a computing node with 24 CPUs. 

Throughout this Section, we will be using two example tasks. The first one, Task 1 (see Listing \ref{list:task1}), taken from \citep{Lee2017b}, consists of finding the probability for a path to exist between two nodes in a graph where edges are probabilistic. This task also appears in \citep{Azzolini2020} and its representation as a PLP is stratified. The second task, Task 2 (see Listing \ref{list:task2}), is the same path finding exercise but now an unknown agent can decide that a node may or may not be included in the graph, and we do not make any assumption as to this agent's behavior: we do not put probabilities on the nodes. We instead represent it as a choice rule\footnote{A choice rule $\{p\}\leftarrow$ is similar to a negative loop $p \leftarrow not q$, $q\leftarrow not p$, making two Answer Sets, one with $p$ and the other without. The difference is that this choice rule doesn't make the atom q appear in the other answer set.} in ASP, meaning that for each total choice there will be multiple answer sets, some in which the node is and some in which it is not. In both tasks there are 6 nodes, edges are generated randomly with random probabilities and we always query the path between node 1 and node 5. Depending on the experiments we vary the number of edges. The listings for the other systems are given in Appendix \ref{ap::t1} and \ref{ap::t2}\footnote{\href{https://github.com/David-Tuc/pasocs\_solve\_exp}{https://github.com/David-Tuc/pasocs\_solve\_exp}}.

\subsection{Convergence}

\begin{figure}[t]
    \centering
    \includegraphics[width=\linewidth]{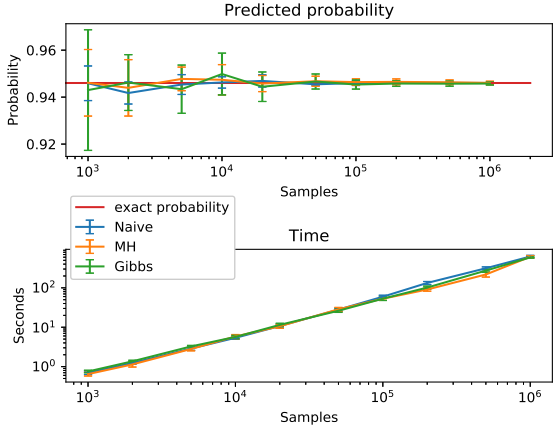}
    \caption{Predicted probability and processing time on task 1 with 20 edges running the three sampling methods for given number of samples. Results are averaged over 10 runs and standard deviation is displayed.}
    \label{fig:task1sampling}
\end{figure}

\begin{figure}[t]
    \centering
    \includegraphics[width=\linewidth]{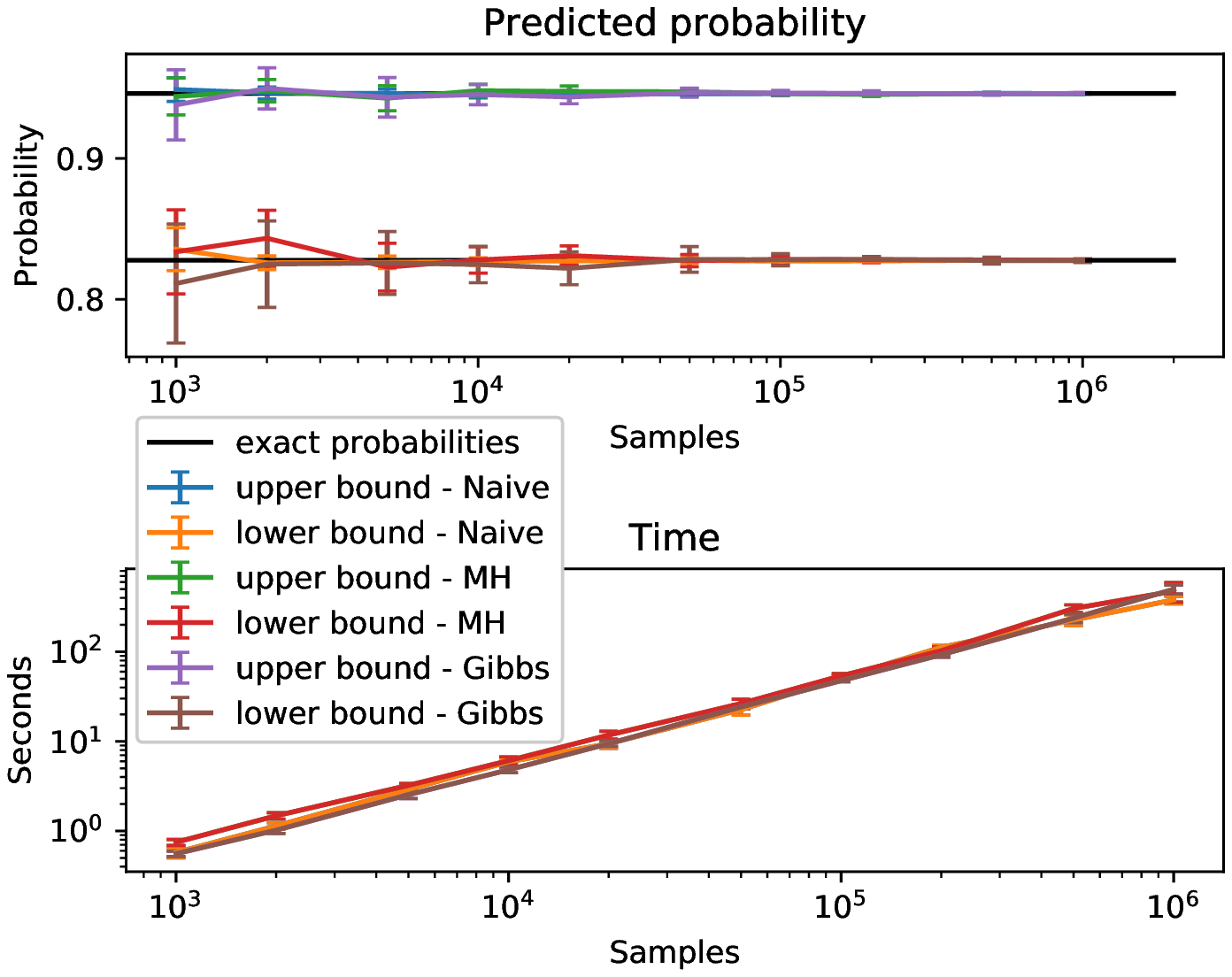}
    \caption{Predicted lower and upper probability and processing time on task 2 with 20 edges running the three sampling methods for given number of samples. Results are averaged over 10 runs and standard deviation is displayed.}
    \label{fig:task2sampling}
\end{figure}

We empirically show the convergence of the sampling methods on stratified and unstratified programs. In the case where the program is stratified (see Figure \ref{fig:task1sampling}) the lower and upper bound probabilities are equal and we obtain a point probability. In the case of an unstratified program, the predicted lower and upper bounds are different (see Figure \ref{fig:task2sampling}). Both of these graphs were obtained by running each sampling method 10 times for each number of samples on the programs with 20 edges.

We see that the standard deviation becomes lower as the number of samples increases and that our sampling methods do converge towards the right estimate given enough samples. Figures \ref{fig:task1sampling} and \ref{fig:task2sampling} also show that the running time is linear with regards to the number of samples and is the same for all three methods (from about 0.57s for $1000$ samples to about 10 minutes for 1 million samples). For both graphs, default parameters were used: $p_{change}=0.3$, $bk=1$ and 100 samples were burned for the MCMC samplings.

\subsection{Sanity Check - stratified programs}

\begin{figure}[t]
\begin{lstlisting}[frame=single,breaklines=true,
label={list:task1},caption={PASOCS Task 1, only two edges are included} ,basicstyle=\small]
0.3576::edge(1, 6).
0.8565::edge(5, 3).
node(1..6).
path(X,Y) :- path(X,Z), path(Z, Y), Y != Z.
path(X,Y) :- node(X), node(Y), edge(X,Y).
#query(path(1,5)).
\end{lstlisting}
\end{figure}

We compare here PASOCS with the following systems: Problog, Cplint, $\mathrm{LP}^{\mathrm{MLN}}$ (LPMLN2ASP \citep{Lee2017b}) and Diff-SAT on task number one. The aim is to show that PASOCS performs as expected and has similar accuracy as other PLP solvers. Now in the context of stratified programs, the semantics of all these systems agree with Sato's Distribution Semantics thus yielding the same results. 

Table \ref{tab:task1exact} shows the running times of exact inference for the given systems. As expected, the running time of PASOCS's exact inference is exponential as it computes each total choice and results are consistent with $\mathrm{LP}^{\mathrm{MLN}}$ which uses the same technique. Problog2 and PITA use knowledge compilation for exact inference which is of course much faster under the condition that the system manages to translate the program. Problog2 was not able in our examples to cope with cycles in the represented graph, which explains the Timeout passed 6 edges.

\begin{table}[t]
    \centering
    \caption{Task 1 time comparison for exact inference (in seconds)}\label{tab:task1exact}
    \begin{tabular}{c c c c c c}
      \toprule 
      \bfseries Edges & \bfseries PASOCS & \bfseries PITA & \bfseries Problog2 & \bfseries $\mathbf{LP}^{\mathbf{MLN}}$\\
      \midrule 
      6 & 0.106 & 0.031 & 1.44 & 1.9 \\
      10 & 0.68 & 0.032 & Timeout & 2,7\\
      16 & 37.34 & 0.031 & Timeout & 64.7\\
      20 & 502 & 0.035 & Timeout & 893\\
      24 & 8748 & 0.037 & Timeout & 15182\\
      \bottomrule 
    \end{tabular}
    \label{Table1}
\end{table}

We also ran approximate inference on Task 1 with 20 edges and report the results in Table \ref{tab:task1samp} using $100000$ samples. All systems find a similar estimate and PASOCS's standard error is only marginally above other systems'. This experiment shows that PASOCS's approximate inference estimate is similar in precision as its competitors on stratified programs, where the Credal semantics matches their respective semantics. If we consider the running times, MCINTYRE is able to perform the sampling in $1.25$ seconds while PASOCS and Diff-SAT take about one minute in average. Problog2 was taking more than an hour and $\mathrm{LP}^{\mathrm{MLN}}$ about two hours to compute the estimate. 

\begin{table}[t]
    \centering
    \caption{Task 1 approximate inference comparison with 100000 samples for 20 edges, averaged over 10 runs. Given with the standard error}\label{tab:task1samp}
    \begin{tabular}{c c c}
      \toprule 
     Systems & \bfseries Probability & \bfseries err \\
      \midrule 
      \bfseries PASOCS &  0.9463 & $1.3e^{-3}$\\ 
      \bfseries MCINTYRE & 0.9459 & $4e^{-4}$\\ 
      \bfseries Problog2 & 0.9464 & $4.84e^{-4}$\\ 
      \bfseries $\mathbf{LP}^{\mathbf{MLN}}$ & 0.9459 & $8.25e^{-4}$\\ 
      \bfseries Diff-SAT & 0.9473 & $1.3e^{-5}$\\ 
      \bottomrule 
    \end{tabular}
\end{table}

\subsection{Comparison - unstratified programs }

\begin{figure}[t]
\begin{lstlisting}[frame=single,breaklines=true,
label={list:task2},caption={PASOCS Task 2, only two edges are included},basicstyle=\small]
0.3576::edge(1, 6).
0.8565::edge(5, 3).
node(1).
{node(2)}.
{node(3)}.
node(4).
node(5).
node(6).
path(X,Y) :- path(X,Z), path(Z, Y), Y != Z.
path(X,Y) :- node(X), node(Y), edge(X,Y).
#query(path(1,5)).
\end{lstlisting}
\end{figure}

While PASOCS is capable of dealing with stratified program, its novelty is that it computes the lower and upper bound probabilities for unstratified programs following the Credal semantics. In this case the only systems we can compare with are $\mathrm{LP}^{\mathrm{MLN}}$ and Diff-SAT. We run each system on Task 2, which includes choice rules. PASOCS will return a lower and upper bound to the probabilities the query can take. 

We report the result of running those systems with $100000$ samples in Table \ref{tab:task2samp}. The lower bound computed by PASOCS has a higher standard error than the upper bound, as it is more reliant on the actual sampled probabilistic facts than the upper bound. We note that the lower and upper bounds computed by PASOCS capture the probabilities computed by the other systems: Diff-SAT's and $\mathrm{LP}^{\mathrm{MLN}}$'s estimate are here approximately their mean value. PASOCS and Diff-SAT take around a minute for that amount of samples while $\mathrm{LP}^{\mathrm{MLN}}$ takes more than five hours.

\begin{table}[t]
    \centering
    \caption{Task 2 approximate inference comparison with 100000 samples for 20 edges, averaged over 10 runs}\label{tab:task2samp}
    \begin{tabular}{c c c c}
      \toprule 
     Systems & \bfseries Probability & \bfseries err & \\ 
      \midrule 
      \multirow{2}{4em}{\bfseries PASOCS} & lower &  0.8282 & $2.4e^{-3}$  \\ 
        & upper &  0.9460 & $1.5e^{-3}$  \\ 
      \bfseries $\mathbf{LP}^{\mathbf{MLN}}$& & 0.8887 & $1.13e^{-3}$ \\ 
      \bfseries Diff-SAT & & 0.8982 & $5.38e^{-4}$ \\ 
      \bottomrule 
    \end{tabular}
\end{table}

In summary, PASOCS is capable of similar if not better performance time-wise as other systems dealing with non-stratified logic programs while allowing more expressiveness in the output probabilities. It returns an estimate of the lower and upper bounds probability as described by the Credal semantics, thus allowing complex programs to be solved without having to make assumptions about unknown probabilities.




\section{Related works}
\label{sec::related_works}

There exist many probabilistic logic programming systems. We have compared with a few in this paper, and we will now detail some of their inner workings and difference with PASOCS. Problog \citep{Fierens2015} and Cplint \citep{Riguzzi2007} work under the distribution semantics \citep{Sato1995}. Problog is a popular solver that uses knowledge compilation to compute the answer to a query (with Problog2), translating the program into a more ordered form which allows for much faster weighted model counting. When this translation succeeds, evaluating the queries is done in linear times. We also compared with its sampling method \citep{Dries2015}. Cplint is implemented as a SWI-Prolog \citep{Colmerauer1990, Wielemaker2012} package that implements both exact and approximate inference. Exact inference is done through PITA \citep{Riguzzi2011a} using knowledge compilation and sampling through MCINTYRE \citep{Riguzzi2011b} using MCMC methods. 

There exist other semantics for PLP which are based on the answer set semantics for logic programs. P-log \citep{Baral2004} is based on a semantics close to the distribution semantics, which assigns a single probability value to queries. It distributes evenly the probability when multiple answer sets correspond to the same total choice. $\mathrm{LP}^{\mathrm{MLN}}$ \citep{Lee2015} is a semantics which takes inspiration both from stable models and Markov logic networks \citep{Richardson2006a}. It annotates logical rules with weights which in turn assign weights to interpretations depending if they are an Answer Set for a subset of the PLP. Weights are then normalised to give a point probability to each interpretation. PrASP \citep{Nickles2015} is a generalisation of the Answer Set semantics which solves an optimization problem defined by a PLP. Its more recent iteration Diff-SAT \citep{Nickles2021} is a more complex system which is capable of computing probabilities of atoms in a PLP through sampling and gradient descent. Like  $\mathrm{LP}^{\mathrm{MLN}}$ and P-log, Diff-SAT assigns point probabilities to queries.

\section{Conclusion}
\label{sec::conclusion}

We presented the new approximate solver PASOCS for probabilistic logic programs under the Credal semantics. It computes the lower and upper bounds of queries given evidence using sampling algorithms. It is aimed at being used with unstratified program but we showed it is similar to other PLP solvers on stratified programs where it agrees with the distribution semantics. 

In the future, PASOCS will aim at solving the equivalent of the Most Probable Explanation and Maximum-A-Posteriori (MAP) tasks under the Credal semantics: cautious explanation and cautious MAP. The system itself will also be kept under constant development to optimize the running time on large clusters, and implement MPI to use multiple computing nodes simultaneously. 

\bibliography{library}
\clearpage
\appendix
\providecommand{\upGamma}{\Gamma}
\providecommand{\uppi}{\pi}
\section{Task 1 Programs}
\label{ap::t1}

We give here the different programs used for task 1, which represent the same graph in the different syntax. We only reproduce here the two first edges and the two first nodes out of six.

\begin{lstlisting}[frame=single,breaklines=true,caption= Problog task 1,basicstyle=\small]
0.3576::edge(1, 6).
0.8565::edge(5, 3).
node(1).
node(2).
path(X,Y) :- path(X,Z), path(Z,Y), Y \== Z.
path(X,Y) :- node(X), node(Y), edge(X,Y).
query(path(1,5)).
\end{lstlisting}

\begin{lstlisting}[frame=single,breaklines=true,caption= MCINTYRE task 1,basicstyle=\small]
:- use_module(library(mcintyre)).
:- mc.
:- begin_lpad.
edge(1, 6):0.3576.
edge(5, 3):0.8565.
node(1).
node(2).
path(Start, End) :- path(Start, End, [Start]).
path(End, End, _).
path(Start, End, Visited) :-edge(Start, Next),\+ memberchk(Next, Visited), path(Next, End,[Next|Visited]).
:- end_lpad.

\end{lstlisting}

\begin{lstlisting}[frame=single,breaklines=true,caption= PITA task 1,basicstyle=\small]
:- use_module(library(pita)).
:- pita.
:- begin_lpad.
edge(1, 6):0.3576.
edge(5, 3):0.8565.
node(1).
node(2).
path(Start, End) :- path(Start, End, [Start]).
path(End, End, _).
path(Start, End, Visited) :- node(Start), node(End), edge(Start, Next),\+ memberchk(Next, Visited), path(Next, End,[Next|Visited]).
:- end_lpad.
\end{lstlisting}

\begin{lstlisting}[frame=single,breaklines=true,caption= $\mathrm{LP}^{\mathrm{MLN}}$ task 1,basicstyle=\small]
@log(0.3576/0.6424000000000001) edge(1, 6).
@log(0.8565/0.14349999999999996) edge(5, 3).
node(1).
node(2).
path(X,Y) :- path(X,Z), path(Z, Y), Y != Z.
path(X,Y) :- node(X), node(Y), edge(X,Y).
\end{lstlisting}

\begin{lstlisting}[frame=single,breaklines=true,caption= Diff-SAT task 1,basicstyle=\small]
_pr_(edge(1, 6), 3576).
{edge(1, 6)}.
_pr_(edge(5, 3), 8565).
{edge(5, 3)}.
node(1).
node(2).
path(X,Y) :- path(X,Z), path(Z, Y), Y != Z.
path(X,Y) :- node(X), node(Y), edge(X,Y).
\end{lstlisting}

\section{Task 2 Programs}
\label{ap::t2}

We give here the different programs used for task 2, which represent the same graph in the different syntax. We only reproduce here the two first edges and make only two nodes capable of flickering (node 2 and 3).

\begin{lstlisting}[frame=single,breaklines=true,caption= $\mathrm{LP}^{\mathrm{MLN}}$ task 2,basicstyle=\small]
@log(0.3576/0.6424000000000001) edge(1, 6).
@log(0.8565/0.14349999999999996) edge(5, 3).
node(1).
{node(2)}.
{node(3)}.
node(4).
node(5).
node(6).
path(X,Y) :- path(X,Z), path(Z, Y), Y != Z.
path(X,Y) :- node(X), node(Y), edge(X,Y).
\end{lstlisting}

\begin{lstlisting}[frame=single,breaklines=true,caption= Diff-SAT task 2,basicstyle=\small]
_pr_(edge(1, 6), 3576).
{edge(1, 6)}.
_pr_(edge(5, 3), 8565).
{edge(5, 3)}.
node(1).
{node(2)}.
{node(3)}.
node(4).
node(5).
node(6).
path(X,Y) :- path(X,Z), path(Z, Y), Y != Z.
path(X,Y) :- node(X), node(Y), edge(X,Y).
\end{lstlisting}




\end{document}